\documentclass[%
reprint,
superscriptaddress,
preprintnumbers,
amsmath,amssymb,
aps,
prd,
]{revtex4-2}

\usepackage[utf8]{inputenc}

\usepackage{mathtools}
\usepackage{amsfonts}
\usepackage{mathrsfs}
\usepackage{bm}
\usepackage{bbold}
\usepackage{slashed}
\usepackage{dsfont}
\usepackage{float}
\usepackage{dcolumn}
\usepackage{amssymb,amsmath}

\usepackage{graphicx}
\usepackage{subcaption}
\usepackage{color}
\usepackage{array}

\usepackage{placeins}
\usepackage{booktabs}
\usepackage{caption}

\usepackage{xspace}
\usepackage{hyperref}
\usepackage[nameinlink]{cleveref}
\usepackage{bookmark}
\usepackage{units}

\def\Eq#1{Eq.~\labelcref{#1}}

\def\eq#1{\labelcref{#1}}
\def\Fig#1{Fig.~\labelcref{#1}}

\def\sec#1{Sec.~\labelcref{#1}}
\def\app#1{App.~\labelcref{#1}}

\setkeys{Gin}{width=0.48\textwidth}
\usepackage{booktabs}
\usepackage{multirow}

\newcolumntype{C}{>{$}c<{$}}
\AtBeginDocument{
	\heavyrulewidth=.08em
	\lightrulewidth=.05em
	\cmidrulewidth=.03em
	\belowrulesep=.65ex
	\belowbottomsep=0pt
	\aboverulesep=.4ex
	\abovetopsep=0pt
	\cmidrulesep=\doublerulesep
	\cmidrulekern=.5em
	\defaultaddspace=.5em
}

\graphicspath{{./figures/}}

\usepackage{xifthen}
\usepackage{xcolor}

\newcommand{\gettitle}{Critical dynamics within the real-time fRG approach}

\hypersetup{
	colorlinks,
	linkcolor={red!75!black},
	citecolor={blue!75!black},
	urlcolor={blue!75!black}, 
	pdftitle={\gettitle},
	pdfauthor={},
	pdfkeywords={}
	{} 
	bookmarksopen=true,
	bookmarksopenlevel=2,
	bookmarksnumbered=true
}

\begin{document}
\title{\gettitle}

\author{Yong-rui Chen}
\affiliation{School of Physics, Dalian University of Technology, Dalian, 116024, P.R. China}	

\author{Yang-yang Tan}
\affiliation{School of Physics, Dalian University of Technology, Dalian, 116024, P.R. China}	

\author{Wei-jie Fu}
\email{wjfu@dlut.edu.cn}
\affiliation{School of Physics, Dalian University of Technology, Dalian, 116024, P.R. China}	
\affiliation{Shanghai Research Center for Theoretical Nuclear Physics, NSFC and Fudan University, Shanghai 200438, China}

\begin{abstract}

The Schwinger-Keldysh functional renormalization group (fRG) developed in \cite{Tan:2021zid} is employed to investigate critical dynamics related to a second-order phase transition. The effective action of model A is expanded to the order of $O(\partial^2)$ in the derivative expansion for the $O(N)$ symmetry.  By solving the fixed-point equations of effective potential and wave function, we obtain static and dynamic critical exponents for different values of the spatial dimension $d$ and the field component number $N$. It is found that one has $z \geq 2$  in the whole range of $2\leq d\leq 4$ for the case of $N=1$, while in the case of $N=4$ the dynamic critical exponent turns to $z < 2$ when the dimension approach towards $d=2$.

\end{abstract}

\maketitle

\section{Introduction}
\label{sec:int}

Non-equilibrium critical dynamics might play a significant role, when quark-gluon plasma (QGP) produced in relativistic heavy ion collisions evolves into the critical region of the critical end point (CEP) in the QCD phase diagram \cite{Son:2004iv, Fu:2022gou, Luo:2022mtp, Arslandok:2023utm}, though recently it has been found the critical region of QCD is extremely small \cite{Braun:2023qak}. This is quite relevant for the search of CEP in experiments under way currently at, e.g., Relativistic Heavy Ion Collider (RHIC) \cite{Luo:2017faz, STAR:2020tga} and other facilities. In the critical region, dynamics is dominated by the massless modes, and the significantly increased correlation length results in the well-known critical slowing down \cite{Rajagopal:1992qz, Berdnikov:1999ph}. On the other hand, the dynamics in the critical region is simplified, since it does not depend on the details of interactions of different systems, but is rather governed by some universal properties, which has been discussed in detail in the seminal paper \cite{Hohenberg:1977ym}.

In early studies of critical dynamics, perturbation techniques are usually adopted, for example the $\epsilon$-expansion \cite{Halperin:1972bwo, Halperin:1974zz, Halperin:1976zz} or the $1/N$ expansion \cite{Ma:1972zz}. In recent years many nonperturbative methods have been utilized. Real-time lattice simulations in classical-statistical field theory are used to calculate spectral functions and critical dynamics \cite{Berges:2009jz, Schlichting:2019tbr, Schweitzer:2020noq, Schweitzer:2021iqk, Florio:2021jlx, Florio:2023kmy}. Real-time correlation functions, spectral functions, dynamic critical exponent, dissipation dynamics, etc., are investigated within the real-time functional renormalization group (fRG) approach \cite{Tan:2021zid, Huelsmann:2020xcy, Roth:2021nrd, Roth:2023wbp, Batini:2023nan}, or based on analytically continued fRG flows \cite{Jung:2021ipc}. Moreover, spectral representations of correlation functions have recently been combined with the fRG and Dyson-Schwinger Equations (DSE), which are now known as the spectral fRG \cite{Braun:2022mgx, Horak:2023hkp} or spectral DSE \cite{Horak:2020eng, Horak:2021pfr, Horak:2022myj, Horak:2022aza}.

In our former work \cite{Tan:2021zid}, we have developed the formalism of fRG formulated on the Schwinger-Keldysh closed time path \cite{Schwinger:1960qe, Keldysh:1964ud}, see also, e.g., \cite{Chou:1984es, Blaizot:2001nr, Berges:2004yj, Sieberer:2015svu} for some relevant reviews about the Schwinger-Keldysh path integral. By the use of the Keldysh rotation, one is able to express the real-time effective action in terms of two different fields: One is referred to as the ``classical'' field and the other ``quantum'' field. The relevant diagram techniques were also devised there \cite{Tan:2021zid}.

In this work we would like to employ the Schwinger-Keldysh fRG in \cite{Tan:2021zid} to study the critical dynamics of a dissipative relaxation model, that is classified as model A in \cite{Hohenberg:1977ym}. We will employ the method of derivative expansion to make truncation for the real-time effective action, which is usually used in the studies of static critical properties, cf. \cite{Balog:2019rrg, DePolsi:2020pjk}. The action is expanded to the order of $O(\partial^2)$ for the $O(N)$ symmetry. Fixed-point equations will be utilized to investigate the critical dynamics of the relaxation model.

This paper is organized as follows: In \sec{sec:model} the dynamic model and its representation in the Schwinger-Keldysh fRG are presented. In \sec{sec:flow} we discuss the flow equations of the effective potential, wave function and the kinetic coefficient. Numerical results are presented and discussed in \sec{sec:num}. In \sec{sec:summary} we show our summary and conclusions. In \app{app:Prop} we show the Feynman rules for the propagators and vertices in the Schwinger-Keldysh fRG. In \app{app:flow-comp} the flow of effective potential in the mesoscopic relaxation model and that in the microscopic theory are compared.

\section{Dynamic model within the real-time fRG approach}
\label{sec:model}

We begin with the dissipative relaxation model with no conservation laws, which is classified as model A in the seminal paper \cite{Hohenberg:1977ym}. The equation of motion for the scalar field of $N$ components $\phi_a$ with $a=0, 1, \cdots N-1$ is described by the Langevin equation, viz.,
\begin{align}
    \frac{\partial \phi_a(x,t)}{\partial t}=&-\Gamma \frac{\delta F[\phi]}{\delta\phi_a}+\eta_a(x,t)\,,\label{eq:Langequ}
\end{align}
with a functional of fields for the free energy
\begin{align}
    F[\phi]=&\int \mathrm{d}^d x\left(\frac{1}{2} Z_\phi(\rho)(\partial_i \phi_a)( \partial_i \phi_a) +V(\rho)-c\sigma\right)\,.\label{eq:free-ener}
\end{align}
where $Z_\phi(\rho)$ and $V(\rho)$ are the field-dependent wave function and effective potential, respectively. The notations $\partial_i=\partial/\partial x^i$, $\rho=\phi^2/2$ with $\phi^2=\phi_a \phi_a$ are used, and summation is assumed for repeated indices. Note that the $O(N)$ symmetry in \Eq{eq:free-ener} is explicitly broken by the last linear term in $\sigma\equiv \phi_{a=0}$, with the breaking strength $c$. In \Eq{eq:Langequ} the diffusion constant $\Gamma$ describes the relaxation rate and the last term denotes the Gaussian white noises with vanishing mean value, i.e., $\langle \eta_a(x,t) \rangle=0$, and nonzero two-point correlations, as follows
\begin{align}
  \langle \eta_a(x,t) \eta_{a'}(x',t')\rangle=&\,2\Gamma \,T\delta(x-x')\delta(t-t')\delta_{aa'}\,,\label{eq:noise-cor}
\end{align}
with the temperature $T$.

In this work, we would like to study the critical dynamics of model A within the functional renormalization group formulated on the Schwinger-Keldysh closed time path. The real-time fRG with the Schwinger-Keldysh path integral and the relevant techniques thereof have been discussed in detail in our former work \cite{Tan:2021zid}, see also \cite{Martin:1973zz, Canet:2006xu}. Following the approach there, one is able to arrive at the renormalization group (RG) scale dependent effective action corresponding to \Eq{eq:Langequ} with \Eq{eq:free-ener}, that is,
\begin{align}
    &\Gamma_k[\phi_c,\phi_q]\nonumber\\[2ex]
=&\int \mathrm{d} t \,\mathrm{d}^d x\,\bigg\{Z_{t,k} \phi_{a,q}  i (\partial_t \phi_{a,c}) +i Z_{\phi,k}(\rho_c) (\partial_i\phi_{a,q}) (\partial_i \phi_{a,c})\nonumber\\[2ex]
&+\frac{i}{4} Z'_{\phi,k}(\rho_c) \phi_{a,q}\phi_{a,c} (\partial_i \phi_{b,c}) (\partial_i \phi_{b,c})+i V'_k(\rho_c)\phi_{a,q}\phi_{a,c}\nonumber\\[2ex]
& -\sqrt{2} i c\sigma_q- 2i\,Z_{t,k} T\phi_{a,q}^2\bigg\}\,,\label{eq:action}
\end{align}
with $\partial_t=\partial/\partial t$, where $\phi_c$ and $\phi_q$ stand for the ``classical'' and ``quantum'' fields, respectively. Note that in \Eq{eq:action} there are derivatives of the wave function and the effective potential, i.e.,
\begin{align}
  Z'_{\phi,k}(\rho_c)=&\frac{\partial Z_{\phi,k}(\rho_c)}{\partial \rho_c}\,,\qquad V'_{k}(\rho_c)=\frac{\partial V_{k}(\rho_c)}{\partial \rho_c}\,,\label{}
\end{align}
with $\rho_c=\phi_c^2/4$. The action here is expanded to the order of $O(\partial^2_i)$ in the derivative expansion. The kinetic coefficient $Z_{t,k}$ in \Eq{eq:action} is related to the relaxation rate in \Eq{eq:Langequ} through $Z_{t,k} \sim 1/\Gamma$. Quantities with suffix $k$ indicate their dependence on the RG scale. From the effective action in the Schwinger-Keldysh field theory in \Eq{eq:action}, one is able to obtain the retarded, advanced, and Keldysh propagators, three- and four-point vertices, etc., which are discussed in detail in \Cref{app:Prop}.

\section{Flow equations}
\label{sec:flow}

%
\begin{figure}[t]
\includegraphics[width=0.4\textwidth]{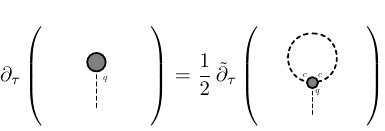}
\caption{Diagrammatic representation of the flow equation for the effective potential, obtained from the one-point correlation function of the effective action with an external leg of $\phi_q$. Here $\tau$ stands for the RG time $\tau=\ln(k/\Lambda)$ with some reference scale $\Lambda$. The partial derivative $\tilde{\partial}_{\tau}$ hits the $k$-dependence only through the regulator in propagators. See \Cref{app:Prop} for more details about the Feynman rule.}\label{fig:Gam1-phiq-equ}
\end{figure}
%
 
The flow equation of effective potential can be obtained from the flow of one-point function as shown in \Fig{fig:Gam1-phiq-equ}. By employing the Feynman rules for the propagators and vertices in \Cref{app:Prop}, one is able to obtain the flow equation of effective potential (first-order derivative w.r.t. the field), to wit,
\begin{align}
  &\partial_\tau V'_k(\rho) \nonumber \\[2ex]
  =&\frac{\nu_d}{2}T k^d \frac{\partial}{\partial \rho}\bigg\{\int_0^1\mathrm{d}x x^{\frac{d}{2}-1}\frac{2-\eta(1-x)}{\big(z_\phi(\rho)-1\big)x+1+\bar m_\sigma^2}\nonumber \\[2ex]
  &+(N-1)\int_0^1\mathrm{d}x x^{\frac{d}{2}-1}\frac{2-\eta(1-x)}{\big(z_\phi(\rho)-1\big)x+1+\bar m_\pi^2}\bigg\}\,,\label{eq:dtV1}
\end{align}
where we have used $\rho$ in replace of $\rho_c$ without ambiguity. The RG time is $\tau=\ln(k/\Lambda)$ with $\Lambda$ being some reference scale, e.g., an ultraviolet cutoff. The angular integral in $d$ dimension gives rise to a constant $\nu_d=1/[(4\pi)^{d/2}\Gamma(d/2)]$, where $\Gamma(d/2)$ is the gamma function. In \Eq{eq:dtV1} one also has
\begin{align}
  z_\phi(\rho)=&\frac{Z_{\phi,k}(\rho)}{Z_{\phi,k}}\,,\label{}
\end{align}
with $Z_{\phi,k}=Z_{\phi,k}(\rho_{0})$, that is field-independent, and $\rho_0$ is usually chosen to be the position of the minimum of potential, i.e., $V'_{k}(\rho_0)=0$. The anomalous dimension reads
\begin{align}
  \eta=&-\frac{\partial_\tau Z_{\phi,k}}{Z_{\phi,k}}\,.\label{eq:eta}
\end{align}
The dimensionless renormalized meson masses in \Eq{eq:dtV1} read
\begin{align}
  \bar m_\sigma^2=&Z_{\phi,k}^{-1}k^{-2}m_\sigma^2\,,\qquad \bar m_\pi^2=Z_{\phi,k}^{-1}k^{-2}m_\pi^2\,,\label{}
\end{align}
where the bare masses are shown in \Eq{eq:ma2}. In \Cref{app:flow-comp} it has been demonstrated that the flow of effective potential in the mesoscopic relaxation model in \Eq{eq:dtV1} corresponds to the high temperature limit of the flow of effective potential in the microscopic Klein-Gordon theory \cite{Tan:2021zid}.

In fact in order to investigate scaling properties of \Eq{eq:dtV1}, it is more convenient to adopt dimensionless renormalized variables, such as
\begin{align}
  \bar \rho=&Z_{\phi,k} T^{-1} k^{2-d}\rho\,,\qquad u(\bar \rho)=T^{-1} k^{-d}V_k(\rho)\,.\label{}
\end{align}
Then, one is left with
\begin{align}
  &\partial_\tau u'(\bar \rho)\nonumber \\[2ex]
  =&(-2+\eta)u'(\bar \rho)+(-2+d+\eta)\bar \rho u^{(2)}(\bar \rho)-\frac{\nu_d}{2}\nonumber \\[2ex]
  &\times \bigg\{\int_0^1\mathrm{d}x\, x^{\frac{d}{2}-1}\big[2-\eta(1-x)\big]\big[z'_\phi(\bar \rho)x+3u^{(2)}(\bar \rho)\nonumber \\[2ex]
  &+2\bar \rho u^{(3)}(\bar \rho)\big]\big[\big(z_\phi(\bar \rho)-1\big)x+1+u'(\bar \rho)+2\bar \rho u^{(2)}(\bar \rho)\big]^{-2}\nonumber \\[2ex]
  &+(N-1) \int_0^1\mathrm{d}x\, x^{\frac{d}{2}-1}\big[2-\eta(1-x)\big]\big[z'_\phi(\bar \rho)x+u^{(2)}(\bar \rho)\big]\nonumber \\[2ex]
  &\times \big[\big(z_\phi(\bar \rho)-1\big)x+1+u'(\bar \rho)\big]^{-2}\bigg\}\,.\label{eq:dtu-prime}
\end{align}

%
\begin{figure}[t]
\includegraphics[width=0.4\textwidth]{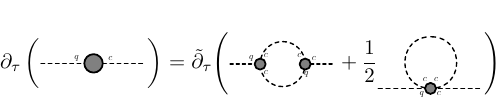}
\caption{Diagrammatic representation of the flow equation for the inverse retarded propagator, see \Eq{eq:Gamm2qc} or \Fig{fig:Gam2-phiqphic}.}\label{fig:Gam2-phiqphic-equ}
\end{figure}
%

%
\begin{figure*}[t]
\includegraphics[width=0.45\textwidth]{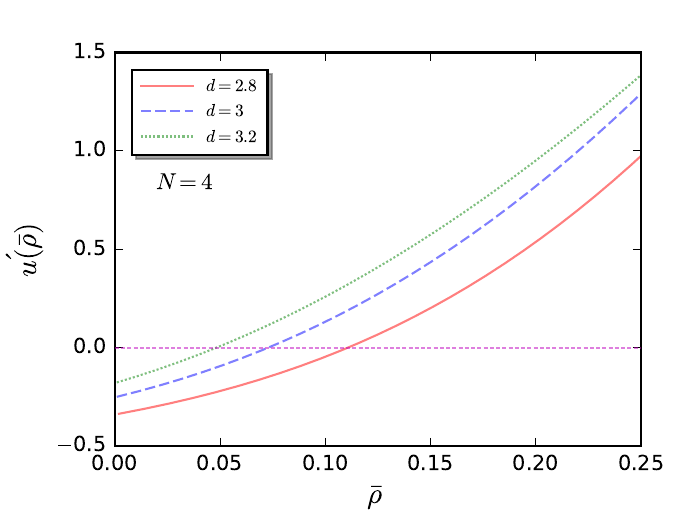}\hspace{0.5cm}
\includegraphics[width=0.45\textwidth]{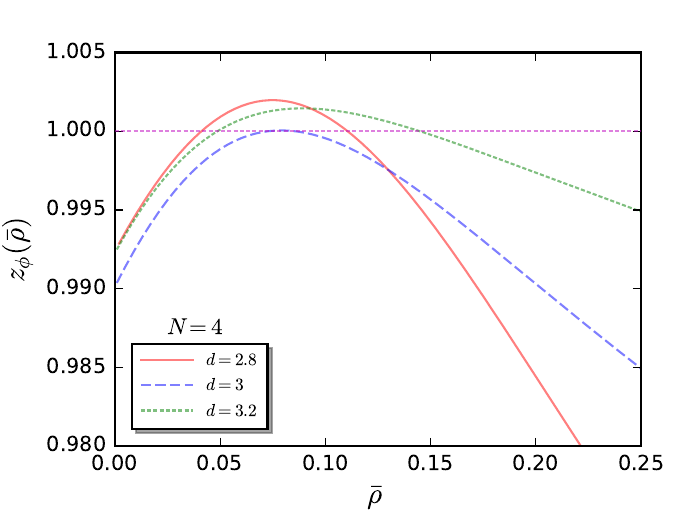}
\caption{Derivative of the effective potential $u^\prime(\bar\rho)$ (left panel) and the wave function $z_\phi(\bar{\rho})$ (right panel) at the Wilson-Fisher fixed point as functions of $\bar\rho$ for the $O(N)$ symmetry with $N=4$. Three different values of dimension $d$ are chosen.}\label{fig:d1u-rho-N4}
\end{figure*}
%

The flows of the wave function renormalization $Z_{\phi,k}(\rho_c)$ and the kinetic coefficient $Z_{t,k}$ in the effective action in \Eq{eq:action} can be extracted from the flow equation of the two-point function, e.g., the inverse retarded propagator in \Eq{eq:Gamm2qc}. The flow of the inverse retarded propagator is shown in \Fig{fig:Gam2-phiqphic-equ}. Inserting the different propagators, three- and four-point vertices in \app{app:Prop} into the flow equation, one is able to close the equations. The computation is straightforward, though a bit tedious. It is obvious from \Eq{eq:Gamm2qc} that the flow of the wave function renormalization can be obtained by performing the projection as follows
\begin{align}
  \partial_{\tau} Z_{\phi,k}(\rho)=&\lim_{\substack{p_0\rightarrow 0\\ \bm{p}\rightarrow 0}}(-i)\frac{\partial}{\partial\bm{p}^2}\frac{\delta^2 \Gamma_{k}[\Phi]}{\delta \phi_{a,q}(-p)\delta \phi_{a,c}(p)}\bigg\vert_{\Phi_{\mathrm{EoM}}}\,.\label{eq:dtZphi}
\end{align}
Note that there is no summation for the index of field component $a$. In the same way, one finds for the kinetic coefficient
\begin{align}
  \partial_{\tau} Z_{t,k}=&\lim_{\substack{p_0\rightarrow 0\\ \bm{p}\rightarrow 0}}\frac{\partial}{\partial p_0}\frac{\delta^2 \Gamma_{k}[\Phi]}{\delta \phi_{a,q}(-p)\delta \phi_{a,c}(p)}\bigg\vert_{\Phi_{\mathrm{EoM}}}\,.\label{eq:dtZt}
\end{align}
In this work the projections in \Eq{eq:dtZphi} and \Eq{eq:dtZt} are made on the pion field, i.e., the field component $a\ne 0$. Then, one arrives at
\begin{align}
  &\partial_\tau z_\phi(\bar{\rho})\nonumber \\[2ex]
  =&\eta z_\phi(\bar{\rho})+(-2+d+\eta)\bar \rho z'_\phi(\bar{\rho})+\frac{2}{d} \bar{\rho}\,\big(z'_\phi(\bar{\rho})\big)^2 \nu_d \nonumber \\[2ex]
  &\times \int_0^1\mathrm{d}x\, x^{\frac{d}{2}}  s(x) \bigg(\frac{1}{L_\pi(x) L^2_\sigma(x)}+\frac{1}{L^2_\pi(x) L_\sigma(x)}\bigg)\nonumber \\[2ex]
  &+4\bar{\rho}z'_\phi(\bar{\rho})u^{(2)}(\bar{\rho}) \nu_d  \int_0^1\mathrm{d}x\, x^{\frac{d}{2} -1} \frac{s(x)}{L_\pi(x)L_\sigma(x)^2}\nonumber \\[2ex]
  &-4\bar{\rho}\big(u^{(2)}(\bar{\rho})\big)^2 \nu_d \int_0^1\mathrm{d}x\, x^{\frac{d}{2} -1} \frac{s(x)}{L^2_\pi(x)L^2_\sigma(x)} \big(\partial_x L_\pi(x)\big) \nonumber\\[2ex]
&+\frac{8}{d} \bar{\rho}\big(u^{(2)}(\bar{\rho})\big)^2\nu_d \int_0^1\mathrm{d}x\, x^{\frac{d}{2}}  \bigg(\frac{1}{L^2_\pi(x)L^3_\sigma(x)}\nonumber\\[2ex]
&+\frac{1}{L^3_\pi(x)L^2_\sigma(x)}\bigg) \big(\partial_x L_\pi(x)\big)^2 s(x)\nonumber\\[2ex]
&-\nu_d \int_0^1\mathrm{d}x\, x^{\frac{d}{2}} \frac{8}{d} \bar{\rho}\big(u^{(2)}(\bar{\rho})\big)^2 \frac{1}{L^2_\pi(x)L^2_\sigma(x)} \big(\partial^2_x L_\pi(x)\big) s(x)\nonumber\\[2ex]
&-\big(z'_\phi(\bar{\rho})+2\bar{\rho}z^{(2)}_\phi(\bar{\rho})\big)\nu_d \int_0^1\mathrm{d}x\, x^{\frac {d}{2} -1}  \frac{1}{L^2_\sigma(x)} s(x)\nonumber\\[2ex]
&-(N-1)z'_\phi(\bar{\rho})\nu_d \int_0^1\mathrm{d}x\, x^{\frac{d}{2} -1}  \frac{1}{L^2_\pi(x)} s(x)\,,\label{eq:dtz}
\end{align}
with
\begin{align}
   s(x)=&\big[2-\eta(1-x)\big]\Theta(1-x)\,,\nonumber\\[2ex]
  L_{\pi}(x)=&z_\phi(\bar\rho)x+(1-x)\Theta(1-x)+u'(\bar{\rho})\,,\nonumber\\[2ex]
  L_{\sigma}(x)=&z_\phi(\bar\rho)x+(1-x)\Theta(1-x)+u'(\bar{\rho}) + 2\bar{\rho}u^{(2)}(\bar{\rho})\,,\label{}
\end{align}
where the flat regulator, cf. \Eq{eq:rB}, is used, and $\Theta(x)$ denotes the Heaviside step function. 

The flow of the kinetic coefficient reads
\begin{align}
  \partial_{\tau} Z_{t,k}=&-Z_{t,k}2\bar{\rho}\big(u^{(2)}(\bar{\rho})\big)^2 \nu_d \int_0^1\mathrm{d}x\, x^{\frac{d}{2} -1}s(x) \nonumber\\[2ex]
&\times \frac{L^2_\pi(x)+4 L_\pi(x)L_\sigma(x)+L^2_\sigma(x)}{L^2_\pi(x) L^2_\sigma(x)\big[L_\pi(x)+L_\sigma(x)\big]^2} \,,\label{eq:dtZt}
\end{align}
which allows us to define the dynamic anomalous dimension as follows
\begin{align}
  \eta_t=&-\frac{\partial_\tau Z_{t,k}}{Z_{t,k}}\,.\label{eq:etat}
\end{align}

The static anomalous dimension in \Eq{eq:eta} can be obtained by evaluating \Eq{eq:dtz} at the minimum of potential $\bar \rho_0$, and one arrives at
\begin{align}
  \eta=&-\partial_\tau z_\phi(\bar \rho_0)+\eta z_\phi(\bar \rho_0)-(\partial_\tau \bar \rho_0) z'_\phi(\bar \rho_0)\,,\label{eq:eta2}
\end{align}
with
\begin{align}
  \partial_\tau \bar \rho_0=&-\frac{\partial_\tau u'(\bar \rho_0)}{u^{(2)}(\bar \rho_0)}\,,\label{}
\end{align}
where the last equation follows from the requirement
\begin{align}
  \frac{d u'(\bar \rho_0)}{d \tau}=&\partial_\tau u'(\bar \rho_0)+(\partial_\tau \bar \rho_0)u^{(2)}(\bar \rho_0)=0\,.\label{}
\end{align}
Note that since the field dependence of the kinetic coefficient is neglected, it is a natural choice to compute its flow in \Eq{eq:dtZt} at the physical point $\bar \rho_0$. With the static anomalous dimension $\eta$ in \Eq{eq:eta} or \Eq{eq:eta2} and dynamic anomalous dimension $\eta_t$ in \Eq{eq:etat}, one is able to calculate the dynamic critical exponent \cite{Hohenberg:1977ym}
\begin{align}
  z=&2-\eta+\eta_t\,\label{eq:z}
\end{align}

It is interesting to find that the flows of the effective potential in \Eq{eq:dtu-prime} and the wave function in \Eq{eq:dtz} do not receive contributions from the dynamical variable $\eta_t$, indicating that the dynamics is decoupled from the static properties in the truncation as shown in the effective action in \Eq{eq:action}. In fact, in more sophisticated truncations, for instance, when the momentum or frequency dependence of the kinetic coefficient $Z_{t,k}$ is taken into account, there is no decoupling any more. Moreover, if the field dependence of the wave function is ignored, the truncation then is reduced to the modified local potential approximation, usually denoted by $\mathrm{LPA}'$. Then the static anomalous dimension reads 
\begin{align}
  \eta=&\frac{8}{d}\frac{1}{2^d\pi^{d/2}\Gamma(d/2)}\frac{\bar \rho_0 \big(u^{(2)}(\bar\rho_0)\big)^2}{\big(1+2\bar \rho_0 u^{(2)}(\bar{\rho}_0)\big)^2}\,,\label{eq:eta-LPApri}
\end{align}
and the dynamic anomalous dimension is given by
\begin{align}
  \eta_t=&\frac{4(2+d-\eta)}{d(2+d)}\frac{1}{2^d\pi^{d/2}\Gamma(d/2)} \bar \rho_0 \big(u^{(2)}(\bar\rho_0)\big)^2 \nonumber\\[2ex]
  &\times\frac{3 +6 \bar \rho_0 u^{(2)}(\bar\rho_0) +2\bar \rho_0^2 \big(u^{(2)}(\bar\rho_0)\big)^2 }{\big(1+\bar\rho_0 u^{(2)}(\bar\rho_0)\big)^2 \big(1+2\bar\rho_0 u^{(2)}(\bar\rho_0)\big)^2}\,.\label{}
\end{align}
If the static anomalous dimension in \Eq{eq:eta-LPApri} is assumed to be vanishing, i.e., $\eta=0$, the truncation then returns to the local potential approximation (LPA).

\section{Numerical results}
\label{sec:num}

%
\begin{figure*}[t]
\includegraphics[width=0.45\textwidth]{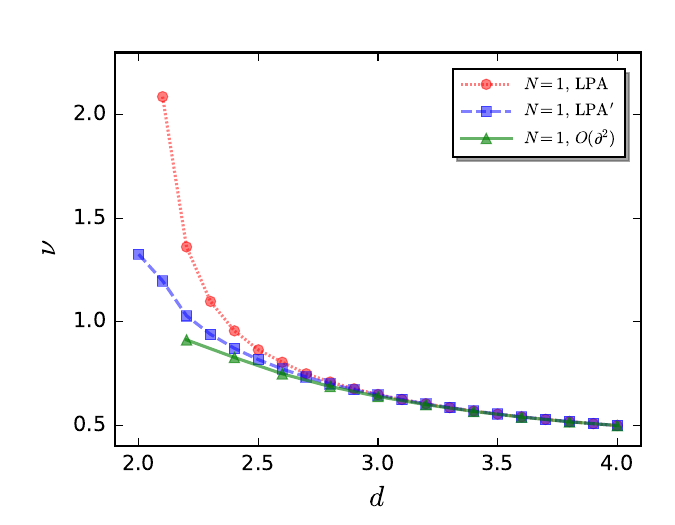}\hspace{0.5cm}
\includegraphics[width=0.45\textwidth]{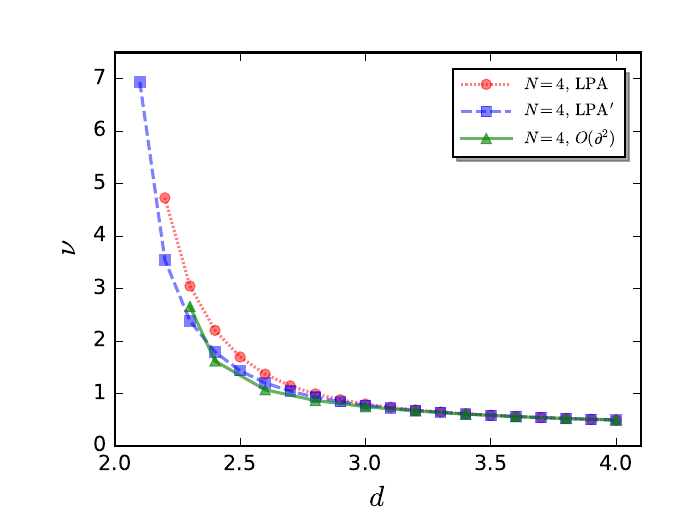}
\caption{Critical exponent $\nu$ as a function of the dimension $d$ for the $O(N)$ symmetry with $N=1$ (left panel) and $N=4$ (right panel). Three different truncations, LPA, $\mathrm{LPA}'$ and derivative expansion up to the order of $O(\partial^2)$ are used, and their respective results are compared.}\label{fig:nu-d}
\end{figure*}
%

%
\begin{figure*}[t]
\includegraphics[width=0.45\textwidth]{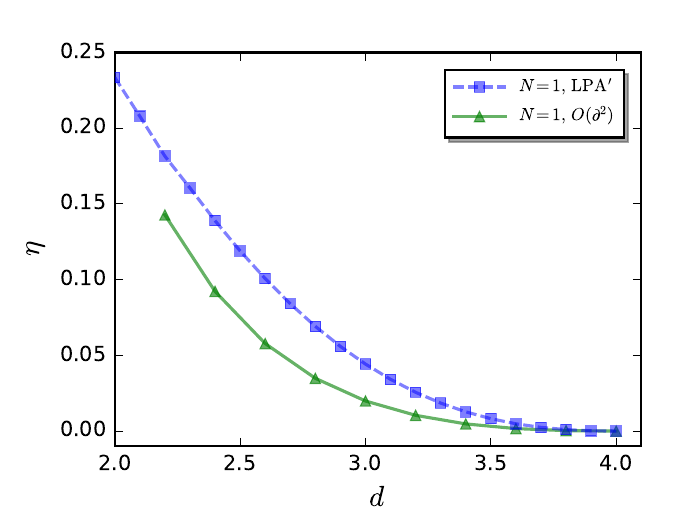}\hspace{0.5cm}
\includegraphics[width=0.45\textwidth]{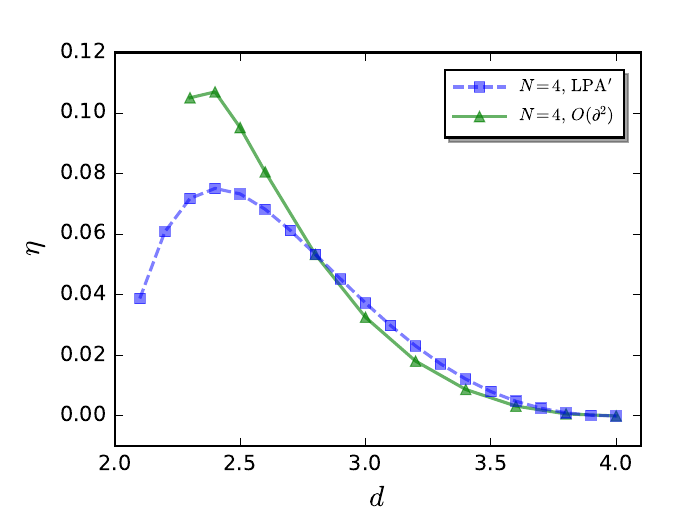}
\caption{Static anomalous dimension $\eta$ as a function of the dimension $d$ for the $O(N)$ symmetry with $N=1$ (left panel) and $N=4$ (right panel). Results in $\mathrm{LPA}'$ are compared with those obtained in derivative expansion in the order of $O(\partial^2)$.}\label{fig:eta-d}
\end{figure*}
%

%
\begin{figure*}[t]
\includegraphics[width=0.45\textwidth]{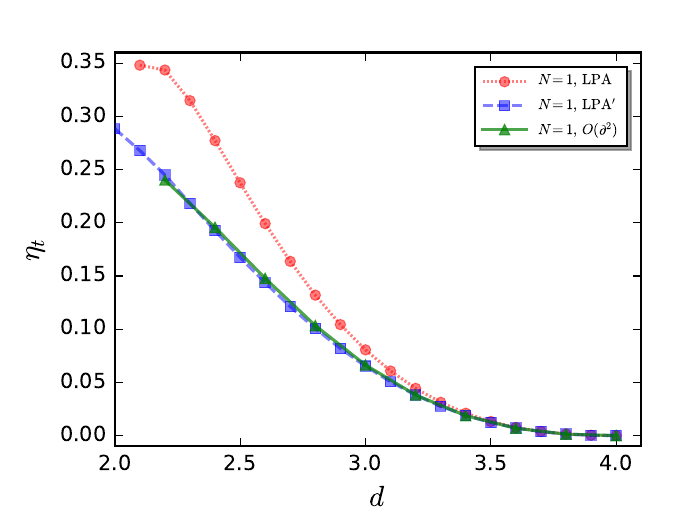}\hspace{0.5cm}
\includegraphics[width=0.45\textwidth]{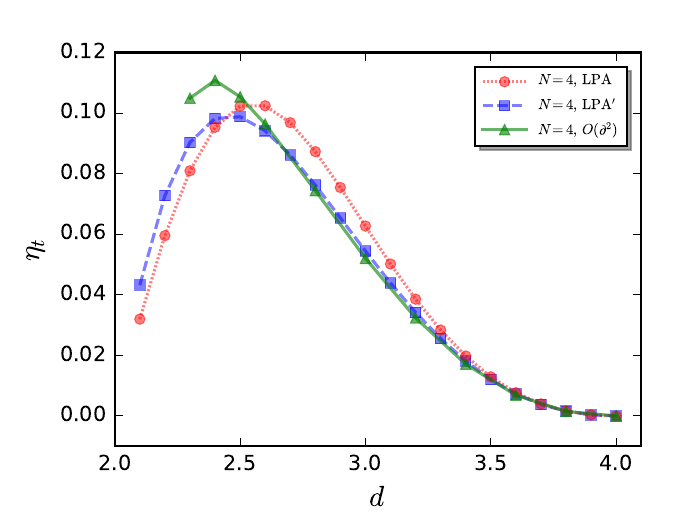}
\caption{Dynamic anomalous dimension $\eta_t$ as a function of the dimension $d$ for the $O(N)$ symmetry with $N=1$ (left panel) and $N=4$ (right panel). Results obtained in three different truncations are compared.}\label{fig:etat-d}
\end{figure*}
%

%
\begin{figure*}[t]
\includegraphics[width=0.45\textwidth]{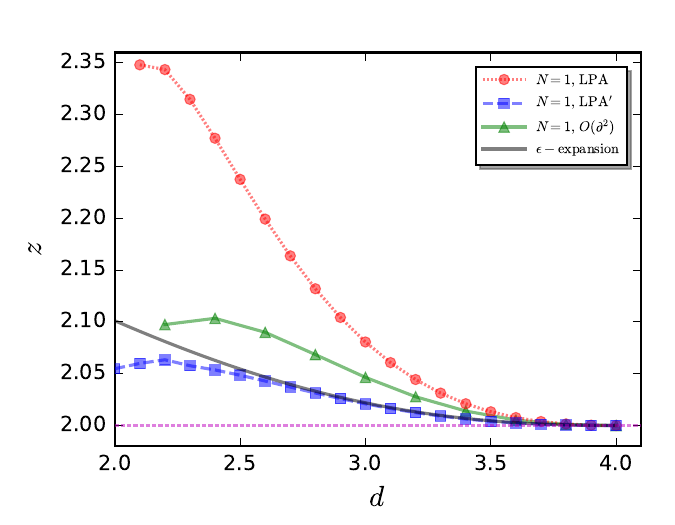}\hspace{0.5cm}
\includegraphics[width=0.45\textwidth]{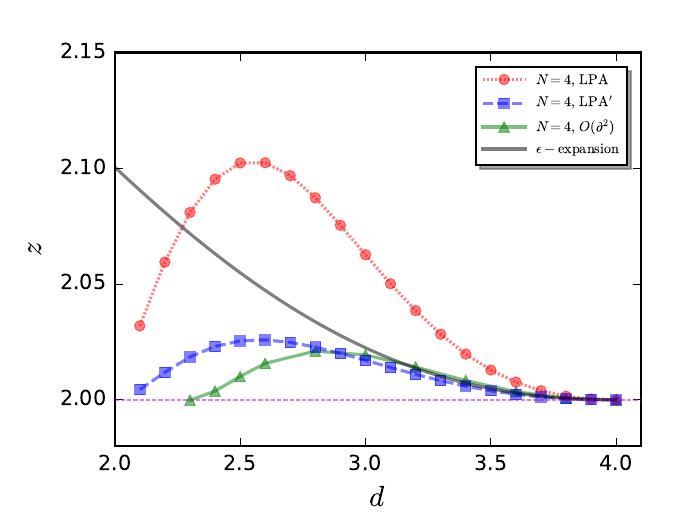}
\caption{Dynamic critical exponent $z$ as a function of the dimension $d$ for the $O(N)$ symmetry with $N=1$ (left panel) and $N=4$ (right panel). Results are obtained in fRG with three different truncations, and they are also compared with the results of three-loop order $\epsilon=4-d$ expansion \cite{Antonov1984}.}\label{fig:z-d}
\end{figure*}
%

In this section we solve fixed-point solutions of the flow equations of the effective potential and wave function numerically, that is $\partial_\tau u'(\bar \rho)=0$ and $\partial_\tau z_\phi(\bar{\rho})=0$. In this work we focus on the Wilson-Fisher fixed point which is characterized by just one relevant eigenvalue of eigenperturbations around the fixed point \cite{Wilson:1971dc, Ma:2020a}, and this relevant eigenvalue is usually denoted by $1/\nu$, where $\nu$ is one of two static critical exponents besides the anomalous dimension $\eta$.

In this work we employ two different numerical methods to solve the fixed-point equations. One is the conventional grid method where the potential $u'(\bar \rho)$ and wave function $z_\phi(\bar{\rho})$ are discretized on a grid of $\bar{\rho}$. The other one is the high-precision direct integral of the fixed-point equation, that is recently proposed in \cite{Tan:2022ksv}, and more details can be found there. We find that these two different numerical methods produce identical results.

In \Fig{fig:d1u-rho-N4} we show the fixed-point solutions of the global potential and wave function. It is found that with the decrease of dimension $d$, the zero crossing point $\bar \rho_0$ of $u'(\bar \rho)$, i.e., $u'(\bar \rho_0)=0$, moves right towards the direction of larger $\bar \rho$. One can also find that with the decrease of dimension $d$, the dependence of the wave function $z_\phi(\bar{\rho})$ on the field $\bar \rho$ becomes stronger, which indicates that when the dimension is small, say $d\lesssim 3$, the field dependence of the wave function should be taken into account. In \Fig{fig:nu-d} we show the critical exponent $\nu$ as a function of the dimension $d$. Obviously, results obtained from the three truncations are convergent when the dimension is $d \gtrsim 3$. Deviations are observed in the region of small $d$, in particular in the vicinity of $d=2$. In fact, the numerical calculations become more and more difficult as the dimension is approaching $d=2$. This is already indicated in the results of the potential in the left panel of \Fig{fig:d1u-rho-N4}. The zero crossing point $\bar \rho_0$ is divergent when one has $d=2$ and $N\geq 2$. Therefore, the calculation of derivative expansion ceases at a value of $d$, where the computation is quite time-consuming.

The results of static anomalous dimension $\eta$ are presented in \Fig{fig:eta-d}. Since one has $\eta=0$ in LPA, only two truncations are compared. One finds that the static anomalous dimension as a function of the dimension is monotonic for $N=1$, whereas there is a non-monotonic dependence in the case of $N=4$. This is closely related to the Mermin-Wagner-Hohenberg theorem \cite{Mermin:1966fe, Hohenberg:1967zz, Coleman:1973ci}, i.e., there is no phase transition in $d=2$ dimension for the $O(N)$ symmetry with $N \geq 2$. In \Fig{fig:etat-d} we show the dynamic anomalous dimension $\eta_t$. Similar with the static one, the dependence of $\eta_t$ on the dimension is monotonic in $N=1$ but non-monotonic for $N \geq 2$. Moreover, one can see $\eta_t=0$ at $d=4$, indicating there is no critical fluctuation or critical slowing down at the Gaussian fixed point.

With the static and dynamic anomalous dimensions, one can obtain the dynamic critical exponent through \Eq{eq:z}. The relevant results are presented in \Fig{fig:z-d}. In the same way, we use three different truncations. Moreover, we also compare with the result of $\epsilon=4-d$ expansion in the order of three loops \cite{Halperin:1972bwo, Folk:2006ve}, which reads
\begin{align}
    z=2+c\eta\,.
\end{align}
with the constant $c$
\begin{align}
    c=0.726(1-0.1885\epsilon)\,,
\end{align}
and the static anomalous dimension in the three-loop order \cite{Wilson:1971vs}
\begin{align}
    \eta=\frac{N+2}{2(N+8)^2}\epsilon^2+\frac{N+2}{2(N+8)^2}\left[\frac{6(3N+14)}{(N+8)^2}-\frac{1}{4}\right]\epsilon^3\,.
\end{align}
One can see that when the dimension is very close to $d=4$, say $d \gtrsim 3.5$, where the $\epsilon$ expansion should work well, our results obtained with the $\mathrm{LPA}'$ or the derivative expansion are comparable with the $\epsilon$ expansion. However, the LPA computation overestimates the dynamic critical exponent apparently, since the static anomalous dimension in LPA is neglected. Furthermore, in the whole range of $2\leq d\leq 4$, our results prefer $z \geq 2$ for the case of $N=1$. The case of $N=4$ is, however, more intricate, and one can see that the dynamic critical exponent calculated with the derivative expansion turns to $z < 2$ when the dimension is below $d=2.5$.

\section{Summary and conclusions}
\label{sec:summary}

In this work, we have used the Schwinger-Keldysh fRG developed in \cite{Tan:2021zid} to study critical dynamics related to a second-order phase transition. As a concrete example, a dissipative relaxation model classified as model A is employed. In the formalism of Schwinger-Keldysh fRG, the RG scale dependent effective action is expressed in terms of two different kinds of fields. One is similar with the physical classical field, i.e. the ``classical'' field, and the other plays the role of fluctuations, called the ``quantum'' field. This formalism of double fields provides us with a very powerful approach to study real-time critical dynamics by employing systematic expansions for the truncation. For example, the derivative expansion can be applied to the sector of classical fields. For the sector of quantum fields, one is able to study the transition from a microscopic theory to a mesoscopic model, e.g., the semiclassical limit of a quantum action, see e.g. \cite{Sieberer:2015svu} for more relevant discussions.

We expand the effective action of $O(N)$ symmetry to the order of $O(\partial^2)$ in the derivative expansion. The flow equations of the effective potential, wave function and the kinetic coefficient are obtained. By solving the fixed-point equations of the dimensionless renormalized potential and wave function, one is able to find the solution of the Wilson-Fisher fixed point and the relevant static and dynamic critical exponents. It is found that both the static anomalous dimension $\eta$ and the dynamic anomalous dimension $\eta_t$ behave as monotonic functions of the spatial dimension $d$ in the range of $2\leq d\leq 4$ in the case of $N=1$, whereas they are both non-monotonic when $N \geq 2$.

The dynamic critical exponent $z$ is obtained as a function of the spatial dimension $d$ for different values of $N$. Our results are also compared with those of $\epsilon$ expansion in the order of three loops. It is found that results obtained from derivative expansion and $\mathrm{LPA}'$ are consistent with that from the $\epsilon$ expansion when the dimension is close to $d=4$, while LPA overestimates the dynamic critical exponent. Furthermore, we find that $z \geq 2$  in the whole range of $2\leq d\leq 4$ for the case of $N=1$, while in the case of $N=4$ the dynamic critical exponent turns to $z < 2$ when the dimension approach towards $d=2$.


\begin{acknowledgments}
We thank Gonzalo De Polsi, Jan M. Pawlowski for discussions. This work is supported by the National Natural Science Foundation of China under Grant Nos. 12175030, 12147101. 
\end{acknowledgments}


\appendix

\section{Propagators and vertices}
\label{app:Prop}

In the functional renormalization group (fRG) approach, quantum and thermal fluctuations of different scales are integrated in successively with the evolution of RG scale \cite{Wetterich:1992yh}. It is a nonperturbative continuum field theory, see, e.g., \cite{Dupuis:2020fhh, Fu:2022gou} for reviews and \cite{Fu:2019hdw, Braun:2020ada, Fu:2021oaw, Chen:2021iuo, Fu:2022uow, Fu:2023lcm} for recent progresses.

\begin{figure}[t]
\begin{align*}
\parbox[c]{0.15\textwidth}{\includegraphics[width=0.15\textwidth]{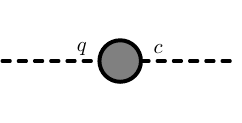}}=i\Gamma^{(2)}_{k,qc}(q)
\end{align*}
\caption{Diagrammatic representation of the two-point function in \Eq{eq:Gamm2qc}.}
\label{fig:Gam2-phiqphic}
\end{figure}

From the effective action in \Eq{eq:action}, one is able to obtain various correlation functions, such as the two-point functions
\begin{align}
  &\Gamma_{k,ba}^{(2)qc}(q',q)=\frac{\delta^2 \Gamma_{k}[\Phi]}{\delta \phi_{b,q}(q')\delta \phi_{a,c}(q)}\bigg\vert_{\Phi_{\mathrm{EoM}}}\nonumber \\[2ex]
  =& \Big(Z_{t,k}\, q_0+i Z_{\phi,k}(\rho_c) \bm{q}^2+i m_{a,k}^2 \Big)\delta_{ab}(2\pi)^{d+1}\delta^{d+1}(q+q')\,,\label{eq:Gamm2qc}
\end{align}
which is depicted in \Fig{fig:Gam2-phiqphic}. Here $\Phi$ denotes all the fields involved in the theory and $\Phi_{\mathrm{EoM}}$ the fields on their respective equations of motion (EoM), i.e., the expectation value of fields. For the classical and quantum fields, one has
\begin{align}
  \phi_{a,q}\vert_{\Phi_{\mathrm{EoM}}}=&0\,,\label{}
\end{align}
and
\begin{align}
  \phi_{a,c}\vert_{\Phi_{\mathrm{EoM}}}=&\left \{\begin{array}{l}
 \phi_{0,c} \qquad a=0\\[3ex]
 0 \qquad \quad a\ne 0
\end{array} \right.,  \label{eq:phic}
\end{align}
which leads to $\rho_c\vert_{\Phi_{\mathrm{EoM}}}=\phi_{0,c}^2/4$. The subscript $\vert_{\Phi_{\mathrm{EoM}}}$ is omitted for brevity, such as in the second line of \eq{eq:Gamm2qc}, if there is no ambiguity. The squared mass in \Eq{eq:Gamm2qc} reads
\begin{align}
  m_{a,k}^2=&\left \{\begin{array}{l}
 m_{a=0,k}^2=m_{\sigma,k}^2=V'_{k}(\rho_c)+2\rho_c V^{(2)}_{k}(\rho_c)\\[3ex]
 m_{a\ne 0,k}^2=m_{\pi,k}^2=V'_{k}(\rho_c)
\end{array} \right.\,.  \label{eq:ma2}
\end{align}
In the same way, one finds for the $\sigma$ and $\pi$ fields
\begin{align}
  \phi_{a,q(c)}=&\left \{\begin{array}{l}
 \sigma_{q(c)} \qquad\;\, a=0\\[3ex]
 \pi_{i, q(c)} \qquad  a=i\ne 0
\end{array} \right.\,.  \label{}
\end{align}

%
%

The relevant infrared regulator reads
\begin{align}
  R_{k,ba}^{qc}(q',q)=& i Z_{\phi,k} \bm{q}^2 r_B\Big(\frac{\bm{q}^2}{k^2}\Big)\delta_{ab}(2\pi)^{d+1}\delta^{d+1}(q+q')\,,\label{}
\end{align}
where $Z_{\phi,k}=Z_{\phi,k}(\rho_{c0})$ is field-independent, and $\rho_{c0}$ is usually chosen to be the position of the minimum of potential, i.e., $V'_{k}(\rho_{c0})=0$. In this work we use the flat regulator \cite{Litim:2000ci, Litim:2001up}, given by
\begin{align}
  r_B(x)&=\Big(\frac{1}{x}-1\Big)\Theta(1-x)\,,  \label{eq:rB}
\end{align}
with the Heaviside step function $\Theta(x)$. Then, one arrives at the inverse retarded propagator \cite{Tan:2021zid}
\begin{align}
  &\mathcal{P}_{k, R}= \Gamma_{k,ba}^{(2)qc}(q)+R_{k,ba}^{qc}(q)\nonumber \\[2ex]
  =&\bigg[Z_{t,k}\, q_0+i Z_{\phi,k}(\rho_c) \bm{q}^2+i Z_{\phi,k}\bm{q}^2 r_B\Big(\frac{\bm{q}^2}{k^2}\Big)+i m_{a,k}^2 \bigg]\delta_{ab}\,.\label{}
\end{align}
where we do not show the delta function of momenta explicitly. Following the same approach, one can also readily obtain the inverse advanced propagator
\begin{align}
  &\mathcal{P}_{k, A}= \Gamma_{k,ba}^{(2)cq}(q)+R_{k,ba}^{cq}(q)\nonumber \\[2ex]
  =&\bigg[-Z_{t,k}\, q_0+i Z_{\phi,k}(\rho_c) \bm{q}^2+i Z_{\phi,k}\bm{q}^2 r_B\Big(\frac{\bm{q}^2}{k^2}\Big)+i m_{a,k}^2 \bigg]\delta_{ab}\,.\label{}
\end{align}
For the $q\,q$ component, one is led to
\begin{align}
  &\mathcal{P}_{k, K}=\Gamma_{k,ba}^{(2)qq}(q',q)=\frac{\delta^2 \Gamma_{k}[\Phi]}{\delta \phi_{b,q}(q')\delta \phi_{a,q}(q)}\bigg\vert_{\Phi_{\mathrm{EoM}}}\nonumber \\[2ex]
  =& -4i Z_{t,k} T\delta_{ab}(2\pi)^{d+1}\delta^{d+1}(q+q')\,,\label{}
\end{align}
Note that there is no regulator in the $q\,q$ component as discussed in \cite{Tan:2021zid}.

Then, we obtain the matrix of inverse propagator 
\begin{align}
  \mathcal{P}_{k}=&\begin{pmatrix} 0 &\mathcal{P}_{k,A}\\[1ex] \mathcal{P}_{k,R} & \mathcal{P}_{k,K} \end{pmatrix}\,,\label{eq:invprop}
\end{align}
and the propagator matrix immediately follows as
\begin{align}
  G_{k}=&\big(\mathcal{P}_{k}\big)^{-1}=\begin{pmatrix} G_{k,K} &G_{k,R}\\[1ex] G_{k,A} & 0 \end{pmatrix}\,,\label{eq:prop}
\end{align}
with
\begin{align}
  G_{k,R}=&(\mathcal{P}_{k,R})^{-1}\,,\qquad G_{k,A}=(\mathcal{P}_{k,A})^{-1}\,, \\[2ex]
 G_{k,K}=&-G_{k,R}\mathcal{P}_{k,K}G_{k,A}\,.\label{eq:GRGAGK}
\end{align}
In short, the retarded, advanced and Keldysh propagators read
\begin{align}
  i G^{R}_{k,ab}=&\langle T_p\phi_{a,c}(x)\phi_{b,q}(y)\rangle\,,\\[2ex]
 i G^{A}_{k,ab}=&\langle T_p\phi_{a,q}(x)\phi_{b,c}(y)\rangle\,,\\[2ex]
 i G^{K}_{k,ab}=&\langle T_p\phi_{a,c}(x)\phi_{b,c}(y)\rangle\,,\label{}
\end{align}
where $T_p$ denotes the time ordering in the closed time path from the positive to negative branch \cite{Tan:2021zid}. In the momentum space by means of Fourier transformation, one arrives at 
\begin{align}
  &i G^{R}_{k,ab}(q)\nonumber \\[2ex]
  =&\frac{i}{Z_{t,k}\, q_0+i Z_{\phi,k}(\rho_c) \bm{q}^2+i Z_{\phi,k}\bm{q}^2 r_B\Big(\frac{\bm{q}^2}{k^2}\Big)+i m_{a,k}^2 }\delta_{ab}\,,\\[2ex]
  &i G^{A}_{k,ab}(q)\nonumber \\[2ex]
  =&\frac{i}{-Z_{t,k}\, q_0+i Z_{\phi,k}(\rho_c) \bm{q}^2+i Z_{\phi,k}\bm{q}^2 r_B\Big(\frac{\bm{q}^2}{k^2}\Big)+i m_{a,k}^2 }\delta_{ab}\,,\\[2ex]
  &i G^{K}_{k,ab}\nonumber \\[2ex]
  =&\frac{4 Z_{t,k}T}{\big(Z_{t,k}\, q_0\big)^2+\Big( Z_{\phi,k}(\rho_c) \bm{q}^2+Z_{\phi,k}\bm{q}^2 r_B\Big(\frac{\bm{q}^2}{k^2}\Big)+ m_{a,k}^2\Big)^2 }\delta_{ab}\,.\label{}
\end{align}
In fact, the retarded propagator is essentially the response function as follows
\begin{align}
  \chi=&i G_{R}\,,\label{}
\end{align}
and the Keldysh propagator is also usually called as the correlation function
\begin{align}
  C=&i G_{K}\,.\label{}
\end{align}
From their expressions above, one can easily show
\begin{align}
  C=&\frac{4T}{q_0}\mathrm{Im}\chi\,,\label{}
\end{align}
which is the fluctuation-dissipation relation.

%
\begin{figure}[t]
\begin{align*}
\parbox[c]{0.15\textwidth}{\includegraphics[width=0.15\textwidth]{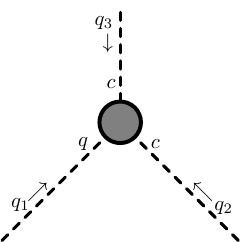}}=i\Gamma^{(3)}_{k,qcc}(q_1,q_2,q_3)
\end{align*}
\caption{Diagrammatic representation of the three-point vertex, where ``$c$'' and ``$q$'' denote the classical and quantum fields, respectively. The momenta $q_1$, $q_2$ and $q_3$ are incoming.}
\label{fig:Gam3-qcc}
\end{figure}
%

%
\begin{figure}[t]
\begin{align*}
\parbox[c]{0.15\textwidth}{\includegraphics[width=0.15\textwidth]{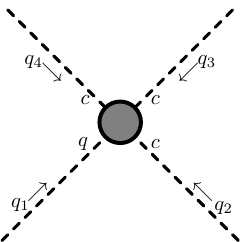}}=i\Gamma^{(4)}_{k,qccc}(q_1,q_2,q_3,q_4)
\end{align*}
\caption{Diagrammatic representation of the four-point vertex.}
\label{fig:Gam4-qccc}
\end{figure}
%

To proceed, we discuss the vertices in the Schwinger-Keldysh field theory, that are relevant in our computation. We begin with the three-point vertex, as follows
\begin{align}
  i\Gamma_{k,a_1a_2a_3}^{(3)qcc}(q_1,q_2,q_3)\equiv& i\frac{\delta^3 \Gamma_{k}[\Phi]}{\delta \phi_{a_1,q}(q_1)\delta \phi_{a_2,c}(q_2)\delta \phi_{a_3,c}(q_3)}\bigg\vert_{\Phi_{\mathrm{EoM}}}\,,\label{}
\end{align}
which is shown diagrammatically in \Fig{fig:Gam3-qcc}. Substituting the effective action in \Eq{eq:action} into the equation above, one obtains after a few calculations
\begin{align}
  &i\Gamma_{k,a_1a_2a_3}^{(3)qcc}(q_1,q_2,q_3)\nonumber \\[2ex]
  =&Z'_{\phi,k}(\rho_c)\rho_c^{1/2}\Big(\bm{q}_1\cdot\bm{q}_2\delta_{a_1a_2}\delta_{a_30}+ \bm{q}_1\cdot\bm{q}_3\delta_{a_1a_3}\delta_{a_20}\nonumber \\[2ex]
  &+\bm{q}_2\cdot\bm{q}_3\delta_{a_2a_3}\delta_{a_1 0}\Big)-\rho_c^{1/2}V_k^{(2)}(\rho_c)\Big(\delta_{a_1a_2}\delta_{a_30}\nonumber \\[2ex]
  &+\delta_{a_1a_3}\delta_{a_20}+\delta_{a_2a_3}\delta_{a_1 0}\Big)-2\rho_c^{3/2}V_k^{(3)}(\rho_c)\delta_{a_10}\delta_{a_20}\delta_{a_30}\,,\label{}
\end{align}
where the delta function of momenta is not shown explicitly. In the same way, the four-point vertex reads
\begin{align}
  &i\Gamma_{k,a_1a_2a_3a_4}^{(4)qccc}(q_1,q_2,q_3,q_4)\nonumber \\[2ex]
  \equiv& i\frac{\delta^4 \Gamma_{k}[\Phi]}{\delta \phi_{a_1,q}(q_1)\delta \phi_{a_2,c}(q_2)\delta \phi_{a_3,c}(q_3)\delta \phi_{a_4,c}(q_4)}\bigg\vert_{\Phi_{\mathrm{EoM}}}\,,\label{}
\end{align}
which is shown in \Fig{fig:Gam4-qccc}. The relevant result can be divided into two parts
\begin{align}
  i\Gamma_{k,a_1a_2a_3a_4}^{(4)qccc}(q_1,q_2,q_3,q_4) =&i\Gamma_{k,\mathrm{I}}^{(4)}+i\Gamma_{k,\mathrm{II}}^{(4)} \,,\label{}
\end{align}
with
\begin{align}
  &i\Gamma_{k,\mathrm{I}}^{(4)} \nonumber \\[2ex]
  =&\frac{1}{2} Z'_{\phi,k}(\rho_c)\Big[\big(\bm{q}_1\cdot\bm{q}_2+\bm{q}_3\cdot\bm{q}_4\big)\delta_{a_1a_2}\delta_{a_3a_4}+ \big(\bm{q}_1\cdot\bm{q}_3\nonumber \\[2ex]
  &+\bm{q}_2\cdot\bm{q}_4\big)\delta_{a_1a_3}\delta_{a_2a_4}+ \big(\bm{q}_1\cdot\bm{q}_4+\bm{q}_2\cdot\bm{q}_3\big)\delta_{a_1a_4}\delta_{a_2a_3}\Big]\nonumber \\[2ex]
  & +\rho_c Z^{(2)}_{\phi,k}(\rho_c)\Big(\bm{q}_1\cdot\bm{q}_2\delta_{a_1a_2}\delta_{a_30}\delta_{a_40}\nonumber \\[2ex]
  &+\bm{q}_1\cdot\bm{q}_3\delta_{a_1a_3}\delta_{a_20}\delta_{a_40}+\bm{q}_1\cdot\bm{q}_4\delta_{a_1a_4}\delta_{a_20}\delta_{a_30}\nonumber \\[2ex]
  &+\bm{q}_2\cdot\bm{q}_3\delta_{a_2a_3}\delta_{a_10}\delta_{a_40}+\bm{q}_2\cdot\bm{q}_4\delta_{a_2a_4}\delta_{a_10}\delta_{a_30}\nonumber \\[2ex]
  &+\bm{q}_3\cdot\bm{q}_4\delta_{a_3a_4}\delta_{a_10}\delta_{a_20}\Big)\,,\label{}
\end{align}
and
\begin{align}
  &i\Gamma_{k,\mathrm{II}}^{(4)} \nonumber \\[2ex]
  =&-\frac{1}{2} V_{k}^{(2)}(\rho_c)\Big(\delta_{a_1a_2}\delta_{a_3a_4}+\delta_{a_1a_3}\delta_{a_2a_4}+\delta_{a_1a_4}\delta_{a_2a_3}\Big)\nonumber \\[2ex]
  & -\rho_cV_{k}^{(3)}(\rho_c)\Big(\delta_{a_1a_2}\delta_{a_30}\delta_{a_40}+\delta_{a_1a_3}\delta_{a_20}\delta_{a_40}\nonumber \\[2ex]
  &+\delta_{a_1a_4}\delta_{a_20}\delta_{a_30}+\delta_{a_2a_3}\delta_{a_10}\delta_{a_40}+\delta_{a_2a_4}\delta_{a_10}\delta_{a_30}\nonumber \\[2ex]
  &+\delta_{a_3a_4}\delta_{a_10}\delta_{a_20}\Big)-2 \rho_c^2 V_{k}^{(4)}(\rho_c)\delta_{a_10}\delta_{a_20}\delta_{a_30}\delta_{a_40}\,.\label{}
\end{align}

%
\begin{figure}[t]
\begin{align*}
\parbox[c]{0.11\textwidth}{\includegraphics[width=0.15\textwidth]{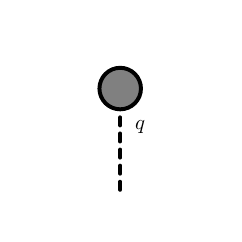}}=i\Gamma^{(1)}_{k,q}
\end{align*}
\caption{Diagrammatic representation of the one-point vertex.}
\label{fig:Gam1-phiq}
\end{figure}
%

Furthermore, in the flow equation of the effective potential in \Fig{fig:Gam1-phiq-equ}, one also needs the one-point vertex as shown in \Fig{fig:Gam1-phiq}, which reads
\begin{align}
  & i\Gamma_{k,a}^{(1)q}(q)\equiv i\frac{\delta \Gamma_{k}[\Phi]}{\delta \phi_{a,q}(q)}\bigg\vert_{\Phi_{\mathrm{EoM}}}\nonumber \\[2ex]
  =&\Big(-2\rho_c^{1/2}V'_k(\rho_c)+\sqrt{2}c \Big)\delta{a0}\,.\label{}
\end{align}
Note that the explicit symmetry breaking $c$ is RG independent.

\section{Comparison of flows of the effective potential in the mesoscopic relaxation model and in the microscopic theory}
\label{app:flow-comp}

In \Eq{eq:dtV1} if the field-dependence of the wave function is ignored, i.e., $z_\phi(\rho)=1$, one can easily simplify the flow. Integrating the field $\rho$ and considering the case of $d=3$, one is led to
\begin{align}
  &\partial_\tau V_k(\rho) \nonumber \\[2ex]
  =&\frac{1}{4 \pi^2}T k^3 \frac{2}{3}\Big(1-\frac{\eta}{5}\Big)\Big(\frac{1}{1+\bar m_\sigma^2}+(N-1)\frac{1}{1+\bar m_\pi^2}\Big)\,.\label{eq:dtV-d3}
\end{align}
In the microscopic Klein-Gordon theory of the $O(N)$ scalar fields, cf. \cite{Tan:2021zid}, the relevant flow of the effective potential is given by
\begin{align}
  &\partial_\tau V_k(\rho) \nonumber \\[2ex]
  =&\frac{1}{4 \pi^2} k^4 \frac{2}{3}\Big(1-\frac{\eta}{5}\Big)\nonumber \\[2ex]
  &\times\bigg\{\frac{1}{(1+\bar m_\sigma^2)^{1/2}}\Big[\frac{1}{2}+\frac{1}{\exp\big(k(1+\bar m_\sigma^2)^{1/2}/T\big)-1}\Big]\nonumber \\[2ex]
  &+\frac{N-1}{(1+\bar m_\pi^2)^{1/2}}\Big[\frac{1}{2}+\frac{1}{\exp\big(k(1+\bar m_\pi^2)^{1/2}/T\big)-1}\Big]\bigg\}\,.\label{eq:dtV-d3-KG}
\end{align}
In the limit of high temperature, one has
\begin{align}
 \frac{1}{\exp\big(k(1+\bar m^2)^{1/2}/T\big)-1}\simeq&\frac{T}{k(1+\bar m^2)^{1/2}}\gg 1\,.\label{}
\end{align}
Then \Eq{eq:dtV-d3-KG} is reduced exactly to \Eq{eq:dtV-d3}.

	
\bibliography{ref-lib}

\end{document}